\documentclass[twocolumn,showpacs,preprintnumbers,amsmath,amssymb]{revtex4}
\usepackage{epsfig}
\newcommand{\bm}[1]{ \mbox{\boldmath $#1$}  }

\begin{document}

\title{The triple alpha reaction rate and the 2$^+$ resonances in $^{12}$C}

\author{R. de Diego and E. Garrido} 
\affiliation{ Instituto de Estructura de la Materia, CSIC, 
Serrano 123, E-28006 Madrid, Spain }

\author{D.V. Fedorov and A.S.~Jensen}
\affiliation{ Department of Physics and Astronomy,
         Aarhus University, DK-8000 Aarhus C, Denmark }

\date{\today}

\begin{abstract}
The triple alpha rate is obtained from the three-body bound and
continuum states computed in a large box. The results from this
genuine full three-body calculation are compared with standard
reference rates obtained by two sequential two-body processes. The
fairly good agreement relies on two different assumptions about the
lowest $2^+$ resonance energy. With the same $2^+$ energy the rates
from the full three-body calculation are smaller than those of the
standard reference. We discuss the rate dependence on the
experimentally unknown $2^+$ energy. Substantial deviations from
previous results appear for temperatures above $3$~GK.
\end{abstract}



\maketitle

\paragraph*{Introduction.}

The triple alpha process is the key reaction that permits to bridge
the $A=5$ and $A=8$ gaps, opening the door to the production of
$^{12}$C in the core of the stars in the red giant phase \cite{sal52}.

The properties of the $\alpha$-$\alpha$-$\alpha$ continuum states are
crucial for the reaction rates, which determine the abundance of
$^{12}$C in the Universe. As a clear example, this is what led Fred
Hoyle to predict the existence of a $0^+$ resonance just above the
triple $\alpha$ threshold \cite{hoy54}.  It is not possible to explain
the observed abundance of $^{12}$C without this so called Hoyle state,
which was confirmed experimentally \cite{coo57} to be at an energy of
0.38 MeV above the three-body threshold. The Hoyle state enhances by
itself the reaction rate by about two orders of magnitude at low
temperatures (below 2 GK), where the rate is dominated by the electric
quadrupole transition from continuum $3\alpha$ 0$^+$ states to
the excited bound $2^+$ state in $^{12}$C.

In type-II supernova explosions dense and hot environments are created. This is the so 
called hot bubble, consisting of alpha-particles and neutrons, with rather uncertain but
relatively high temperature reaching several GK \cite{arn99}. These environments are the 
suggested place for the rapid-neutron process 
\cite{mey92}. Although the importance of the triple alpha contribution as compared to other 
reactions in this temperature range ($T \gtrsim 2$ GK) has been questioned \cite{efr96}, it has 
also been suggested that changes in this rate imply changes in estimates of the relative 
amounts of elements formed during the supernovae explosion, and therefore in estimates of the
rate at which heavy elements are distributed through the Universe \cite{eid05}.
In \cite{fyn05} the estimated reaction rate for the triple alpha reaction in this temperature 
range was found to be about an order of magnitude lower than the one given in \cite{ang99}, which 
would lead to a mass fraction of $^{56}$Ni about two to three times smaller.

At temperatures above 3 GK, the reaction rate for the triple alpha
reaction is dominated by the electric quadrupole $2^+ \rightarrow
0^+_1$ transition \cite{ang99} 
(we follow the notation where $J^\pi_i$ represents the $i^{th}$ state in the spectrun
with angular momentum $J$ and parity $\pi$. In the
particular case of $^{12}$C the states 0$^+_1$ and 2$^+_1$ are the only ones bound).
The reaction rate in this region is
therefore to a large extent determined by the properties of the $2^+$
continuum states, and in particular by the possible existence of a
$2^+_2$ resonance at a relatively low energy. The properties of such a
state is still an open problem.  The reaction rates given in
\cite{ang99,des87} were computed assuming that $^{12}$C has a $2^+_2$
resonance at 1.75 MeV. However, in \cite{fyn05} no evidence was found
concerning the existence of such $2^+$ resonance. On the contrary, in
\cite{fre09} an experimental energy of about 2.3 MeV is derived. This 
value is substantially higher than the one used in \cite{ang99,des87},
and it could substantially modify their computed results. Also, in
\cite{alv07}, where a three-body calculation is employed to obtain the
resonances as poles of the $S$-matrix, a 2$^+_2$ resonance has been
found at an energy of 1.38 MeV. However, this energy is found with the
same (adjustable) three-body interaction that reproduces the energy of
the bound $2^+_1$ state.

The role of the 2$^+_2$ resonance in $^{12}$C in the triple alpha
reaction rate requires detailed investigation. We shall employ the
same three-body method as in \cite{die10}, where no assumption is made
about the capture mechanism (sequential or direct), and in contrast to
the method described in \cite{ang99}, where a sequential capture
process is assumed.  The standard procedure and the almost canonical
results of \cite{ang99} is furthermore tested by comparison.

\paragraph*{Full three-body formulation.}

The first of the methods used in this work is the three-body
calculation described in \cite{die10}. Let us consider the radiative
capture process $a+b+c\rightarrow A+\gamma$, where $A$ is a bound
system made of particles $a$, $b$, and $c$ with separation energy
$B_A$. The corresponding reaction rate $R_{abc}(E)$ is given by
\cite{fow67,die10}
\begin{equation}  
R_{abc}(E)=\frac{\hbar^3}{c^2} \frac{8\pi}{(\mu_x \mu_y)^{3/2}} 
\left( \frac{E_\gamma}{E} \right)^2 \frac{2 g_A}{g_a g_b g_c} 
\sigma_\gamma(E_\gamma)
\label{eq1} 
\end{equation}
where $E=E_\gamma+B_A$ is the initial three-body kinetic energy,
$E_\gamma$ is the photon energy, $\sigma_\gamma(E_\gamma)$ is the
photo dissociation cross section of the $A$ nucleus, $c$ is the
velocity of light, $g_i$ is the spin degeneracy of states of particle
$i=a,b,c,A$, and $\mu_x$ and $\mu_y$ are the reduced masses of the
systems related to the Jacobi coordinates, $(\bm{x},\bm{y})$, for the
three-body system \cite{nie01}.

The photo dissociation cross section for the inverse process
$A+\gamma\rightarrow a+b+c$ can be expanded into electric and magnetic
multipoles. In particular, the electric multipole contribution
of order $\lambda$ has the form \cite{die10}:
\begin{equation} 
\sigma^{(\lambda)}_\gamma(E_\gamma)=\frac{(2 \pi)^3 (\lambda+1)}{\lambda [(2 \lambda+1)!!]^2}  
\left(\frac{E_\gamma}{\hbar c}\right)^{2 \lambda-1} \frac{d{\cal B}}{dE} \;, 
\label{eq2} 
\end{equation}
where the strength function ${\cal B}$ is
\begin{equation} 
   {\cal B}(E\lambda,n_0J_0 \rightarrow nJ) = \sum_{\mu M} 
  |\langle nJM|O_{\mu}^{\lambda}|n_0J_0M_0\rangle|^2, 
\label{eq3} 
\end{equation}
where $J_0$, $J$ and $M_0$, $M$ are the total angular momenta and
their projections of the initial and final states, and all other
quantum numbers are collected into $n_0$ and $n$.  The electric
multipole operator is given by:
\begin{eqnarray}  
 O_{\mu}^{\lambda} = \sum_{i=1}^3 z_i |\bm{r}_i - \bm{R}|^{\lambda} 
 Y_{\lambda, \mu}(\Omega_{y_i}) \;, 
\label{eq4}
\end{eqnarray}
where $i$ runs over the three particles of charges $z_i$ , and where we
neglect contributions from intrinsic transitions within each of the
three constituents \cite{rom08}.

Finally the energy averaged reaction rate is obtained as a function of
the temperature by using the Maxwell-Boltzmann distribution as
weighting function. For three alpha particles we obtain \cite{die10}:
\begin{eqnarray} 
\lefteqn{  \hspace*{-1cm} 
\langle R_{\alpha\alpha\alpha}(E) \rangle= \frac{\hbar^3}{c^2} 
\frac{48\pi}{(\mu_{\alpha\alpha} \mu_{\alpha ^8\mbox{\tiny Be}})^{3/2}} 
(2J+1) e^{-\frac{B}{k_B T}}\times } \nonumber \\ && \times 
\frac{1}{(k_B T)^3} \int_{|B|}^{\infty} E_\gamma^2 
\sigma^{(\lambda)}_\gamma(E_\gamma)e^{-\frac{E_\gamma}{k_B T}} dE_\gamma\;,
\label{eq5} 
\end{eqnarray}
where $k_B$ is the Boltzmann constant.

The strength function ${\cal B}$ is computed by genuine three-body
calculations of both the bound final state, $|n_0J_0M_0\rangle$, and
the continuum initial states, $|nJM\rangle$.  We use the hyperspherical
adiabatic expansion method described in \cite{nie01}. The
$\alpha$-$\alpha$ interaction is given in \cite{alv08}. The
basic procedure is computation of three-body states of given angular
momentum and parity confined by box boundary conditions
\cite{die07}. In this way the continuum spectrum is discretized.  The
strength functions are then obtained for each discrete continuum state
according to Eq.(\ref{eq3}). The distribution $d{\cal B}/dE$ is built
by use of the finite energy interval approximation, where the energy
range is divided into bins, and all the discrete values of ${\cal B}$
falling into a given bin are added. Afterwards the points are
connected by spline operations and the expressions (\ref{eq2}) and
(\ref{eq5}) are computed.

\paragraph*{Sequential process.}
The NACRE results given in \cite{ang99} are usually taken as the
reference for the reaction rates in the triple alpha process. In this
work the process is assumed to proceed in a sequentially two-step
process. In the first step one $\alpha$-particle captures another one
to produce $^8$Be in the ground $0^+$ resonant state. In the second
step, $^8$Be is able to capture (before decaying) a third
$\alpha$-particle, populate a $^{12}$C resonance, and then decay by
photo emission to one of the bound states of $^{12}$C. The reaction
rate for such a two-step process is given by the rate for the capture
of an $\alpha$-particle by $^8$Be ($\langle R_{\alpha\mbox{\tiny
$^8$Be}}(E'',E^\prime) \rangle$) weighted with the rate for formation
of $^8$Be \cite{ang99}:
\begin{eqnarray}
\lefteqn{
\langle R_{\alpha\alpha\alpha}(E^\prime)\rangle= 3 \frac{8 \pi\hbar}{\mu_{\alpha\alpha}^2}
\left(\frac{\mu_{\alpha\alpha}}{2\pi k_BT} \right)^{3/2}  } \label{eq6} \\ & & 
\int_0^\infty \frac{\sigma_{\alpha\alpha}(E'')}{\Gamma_\alpha(^8\mbox{Be},E'')}e^{-E''/k_BT} \langle 
R_{\alpha\mbox{\tiny $^8$Be}}(E'',E^\prime) \rangle E'' dE'', \nonumber
\end{eqnarray}
where $E''$ and $E^\prime$ are the relative energy between the two
$\alpha$-particles forming $^8$Be and the energy of the third
$\alpha$-particle relative to the center of mass of the first two,
respectively. The temperature is $T$ and $\mu_{\alpha\alpha}$ is the
reduced mass of the two-alpha system. The elastic $\alpha-\alpha$
cross section is given by:
\begin{equation}
\sigma_{\alpha\alpha}(E'')=\frac{2\pi}{\kappa^2}\frac{\Gamma_\alpha(^8\mbox{Be},E'')^2}{(E''-E_r)^2+(\Gamma_\alpha(^8\mbox{Be},E''))^2/4},
\label{eq7}
\end{equation}
where $\kappa^2=2\mu_{\alpha\alpha}E''/\hbar^2$, $E_r$ is the $^8$Be
resonance energy, and the width $\Gamma_\alpha(^8\mbox{Be},E'')$ has
the form \cite{ang99}:
\begin{equation}
\Gamma_\alpha(^8\mbox{Be},E'')= \Gamma_\alpha \frac{P_\ell(E'')}{P_\ell(E_r)},
\end{equation}
where $\Gamma_\alpha$ is the width of the resonance, $P_\ell$ is the penetration factor, and $\ell$ is
the relative orbital angular momentum between the two $\alpha$-particles.

In our case, the experimental energy of the 0$^+$ resonance in $^8$Be
is $E_r=0.09189$ MeV above threshold, with a width of
$\Gamma_\alpha=6.8\pm 1.7 \cdot 10^{-6}$ MeV. For such a narrow
resonance we have that $\Gamma_\alpha(^8\mbox{Be},E'')\approx
\Gamma_\alpha$, and for our purpose we can safely replace
Eq.(\ref{eq7}) by:
\begin{equation}
\frac{\sigma_{\alpha\alpha}(E'')}{\Gamma_\alpha(^8\mbox{Be},E'')}=\frac{4\pi^2}{\kappa^2} \delta(E''-E_r),
\label{eq9}
\end{equation}
from which Eq.(\ref{eq6}) becomes:
\begin{eqnarray}
\lefteqn{\hspace*{-1.5cm}
\langle R_{\alpha\alpha\alpha}(E^\prime)\rangle= 3  \frac{8 \pi\hbar}{\mu_{\alpha\alpha}^2}
\left(\frac{\mu_{\alpha\alpha}}{2\pi k_BT} \right)^{3/2}  }  \label{eq10} \\ & & 
\times \frac{4\pi^2}{\kappa^2} E_r e^{-E_r/K_BT} \langle R_{\alpha\mbox{\tiny $^8$Be}}(E^\prime) \rangle \nonumber
\end{eqnarray}

Finally, from Ref.\cite{ang99} we have:
\begin{eqnarray}
\lefteqn{\hspace*{-1.5cm}
\langle R_{\alpha\mbox{\tiny $^8$Be}}(E^\prime)\rangle= \frac{8 \pi            }{\mu_{\alpha\mbox{\tiny $^8$Be}}^2}
\left(\frac{\mu_{\alpha\mbox{\tiny $^8$Be}}}{2\pi k_BT} \right)^{3/2}  } \nonumber \\ & & 
\times \int_0^\infty \sigma_{\alpha\mbox{\tiny $^8$Be}}(E^\prime)e^{-E^\prime/k_BT} E^\prime dE^\prime, \label{eq11}
\end{eqnarray}
where
\begin{eqnarray}
\lefteqn{\hspace*{-1.5cm}
\sigma_{\alpha\mbox{\tiny $^8$Be}}(E^\prime)=\sum_{J=0,2}(2J+1) \frac{\pi \hbar^2}{2\mu_{\alpha\mbox{\tiny $^8$Be}}E^\prime}
 } \nonumber \\ & & \times
\frac{\Gamma_\alpha(^{12}C^J,E^\prime) \Gamma_\gamma(^{12}C^J,E^\prime)}
     {(E^\prime-E_r^J)^2+0.25 \Gamma(^{12}C^J,E^\prime)^2}
\label{eq12}
\end{eqnarray}
where $E_r^J$ is the $^{12}$C resonance energy with angular momentum $J$, and
\begin{equation}
\Gamma(^{12}C^J,E^\prime)=\Gamma_\alpha(^{12}C^J,E^\prime)+\Gamma_\gamma(^{12}C^J,E^\prime)
\label{eq13}
\end{equation}
\begin{equation}
\Gamma_\alpha(^{12}C^J,E^\prime)=\Gamma_\alpha(^{12}C^J) \frac{P_\ell(E^\prime)}{P_\ell(E_r^J)}
\label{eq14}
\end{equation}
\begin{equation}
\Gamma_\gamma(^{12}C^J,E^\prime)=\Gamma_\gamma(^{12}C^J)\frac{(E_T^J+E^\prime)^5}{(E_T^J+E_r^J)^5}
\label{eq15}
\end{equation}

\begin{table}
\begin{tabular}{c|cccc} 
      & $E_r^J$ & $\Gamma_\alpha(^{12}\mbox{C}^J)$ &  $\Gamma_\gamma(^{12}\mbox{C}^J)$ & $E_T^J$ \\
 \hline
$J=0$ & 0.2877        &  $8.3 \cdot 10^{-6}$ & $3.7\cdot 10^{-9}$  &   2.928   \\
$J=2$ & 1.75          &  0.56               & $ 0.2\cdot 10^{-6}$  &   7.367   
\end{tabular}
\caption{Values of the parameters entering in Eqs.(\ref{eq12}) to (\ref{eq15}) taken from Ref.\cite{ang99}.
They are all given in MeV.}
\label{tab1}
\end{table}
The parameters needed to compute Eq.(\ref{eq12}), and therefore (\ref{eq11}) and (\ref{eq10}), are taken 
from \cite{ang99} and they are collected in table~\ref{tab1}.

Given a transition between $^{12}$C continuum states with angular
momentum $J$ and a bound $^{12}$C state with binding energy $B_J$, it
is now possible to extract the expression for $d{\cal B}/dE$
effectively used in Ref.\cite{ang99}. This can be made by inserting
Eq.(\ref{eq11}) into (\ref{eq10}), and Eq.(\ref{eq2}) with $\lambda$=2
into Eq.(\ref{eq5}). By comparison of those expressions we extract:
\begin{equation}
\frac{d{\cal B}}{dE}=\frac{1}{\alpha} \frac{375}{8\pi^2} \frac{(\hbar c)^4}{(E+|B_J|)^5}
\frac{\Gamma_\alpha(^{12}C^J,E^\prime) \Gamma_\gamma(^{12}C^J,E^\prime)}
     {(E^\prime-E_r^J)^2+0.25 \Gamma(^{12}C^J,E^\prime)^2},
\label{eq16}
\end{equation}
where $E=E_r+E^\prime$ is the three-body kinetic energy and $\alpha$ is the fine structure constant.

\paragraph*{Comparing the methods.}

\begin{figure}[t]
\psfig{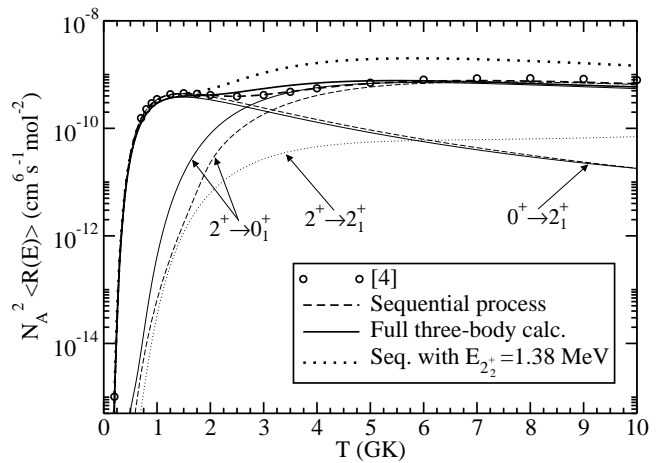}
\caption{Reaction rate for the triple $\alpha$ process with the full three-body calculation (thick
solid line) and the sequential approach (thick dashed line) as described in the text. The corresponding
contributions from the $0^+\rightarrow 2^+_1$ and $2^+\rightarrow 0^+_1$ transitions are given by the
thin curves ($2^+_1$ and 0$^+_1$ refer to the lowest $2^+$ and 0$^+$ states in the $^{12}$C 
spectrum, which are the only ones bound). 
The thin dotted line is the contribution from the $2^+\rightarrow 2^+_1$ in the full three-body 
calculation. The open circles correspond the rate given in \cite{ang99}. The thick dotted curve is
a sequential calculation but using the same energy and width for the 2$^+_2$ resonance as in the
three-body calculation.}
\label{fig1}
\end{figure}

In \cite{die10} the full three-body method was used to compute the reaction rate for the triple $\alpha$ process.
The $\alpha$-$\alpha$ interaction in \cite{alv08} was used. This leads to the $^{12}$C spectrum given
in \cite{alv07}, which results in good agreement with the experimental data for about 13 different
states. In particular, concerning the 2$^+$ resonances, the computed $^{12}$C spectrum includes a 
low lying one at 1.38 MeV (above the three-body threshold) with a computed width of 0.13 MeV, and a second 
one at 4.4 MeV with a width of about 1 MeV. Experimentally only a resonance at 3.88 MeV is fully established. 
In Ref.\cite{alv07} the resonances are obtained as poles of the $S$-matrix by use of the complex scaling method
\cite{fed03}. With this method the resonances behave asymptotically as bound states with complex energy,
which automatically determines the resonance energy and the corresponding width. The resonance width is then not 
a parameter but an output of the calculation. The three-body calculation, as described in \cite{die10}, 
gives rise to the triple $\alpha$ reaction rate shown by the thick solid line in
Fig.\ref{fig1}. The contributions coming from the $0^+\rightarrow 2^+_1$ and $2^+\rightarrow 0^+_1$ transitions
are shown by the corresponding thin solid lines.  The agreement between the total reaction
rate and the result given in \cite{ang99} (open circles in the figure) is reasonably good. For completeness, we
show in the figure the contribution from the   $2^+\rightarrow 2^+_1$ transition (thin dotted curve). 
This contribution is very small and could actually be neglected.

When the sequential approach is used, and the reaction rate is
obtained from Eq.(\ref{eq10}), we get the thick dashed curve in the
figure. The sequential method described in the present work and in
\cite{ang99} are identical except for the approximation in
Eq.(\ref{eq9}). The parameters used in the calculations
(table~\ref{tab1}) are also the same in both cases. The good agreement
between the thick dashed curve and the circles in Fig.\ref{fig1} shows
that the approximation in Eq.(\ref{eq9}) is accurate. Only a small
difference is found for high temperatures since the contribution from
the 3$^-_1$ resonance in $^{12}$C is included in \cite{ang99}, but
completely neglected in our calculation. Again, the corresponding thin
dashed lines show the contributions to the total rate from the
$0^+\rightarrow 2^+_1$ and $2^+\rightarrow 0^+_1$ transitions.

As seen in Fig.\ref{fig1}, the contribution from the $0^+\rightarrow
2^+_1$ transition is very similar in the full three-body calculation
and the sequential approximation. This is due to the narrow 0$^+_2$
Hoyle three-body state in $^{12}$C which heavily dominates the full
calculation. It has a large strength corresponding to strong
population and subsequent decay into the bound 2$^+_1$ state plus a
photon. This process is precisely as assumed in the sequential
description, and in fact the reason for its success.

However, the lowest $2^+_2$ resonance in $^{12}$C (both the one obtained
in the three-body calculation and the one used in \cite{ang99}) is
rather wide, which means that the continuum non-resonant three-body
2$^+$ states can contribute significantly to the $2^+\rightarrow
0^+_1$ transition. In fact, as seen in the figure, the three-body and
the sequential calculation differ for this contribution, especially at
low temperatures. Due to this discrepancy the two computed total reaction 
rates do not fully agree for temperatures from about 2 to 5 GK. When the 
temperature increases the thin solid and thin dashed curves in Fig.\ref{fig1}
become closer and closer, in such a way that for temperatures higher then
5 GK the total reaction rate is similar in both calculations.

\begin{figure}[ht]
\psfig{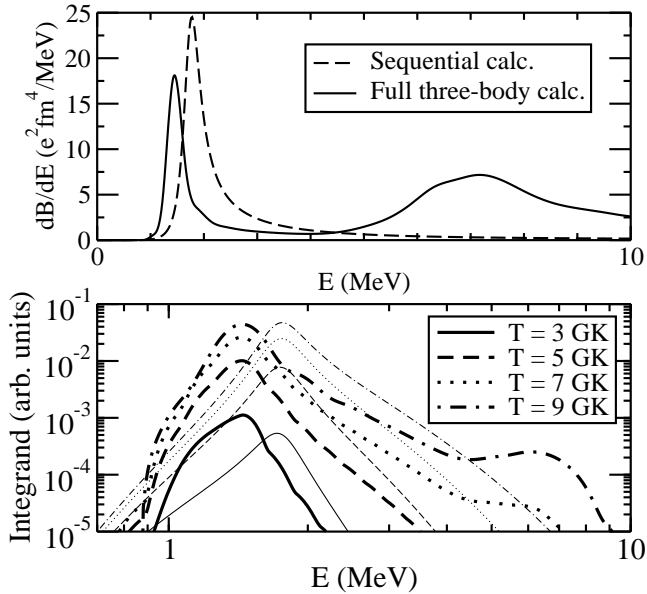}
\caption{Upper part: the $d{\cal B}/dE$ function for the $2^+\rightarrow 0^+_1$ transition for the sequential 
approach as given in Eq.(\ref{eq16}) (dashed curve) and the three-body calculation (solid curve). Lower part:
Integrands of Eq.(\ref{eq5}) (thick curves) and Eq.(\ref{eq10}) (thin curves) in arbitrary units as a function 
of the kinetic energy of the three $\alpha$'s for $T$=3, 5, 7, and 9 GK.
The thick lines and thin lines correspond to the full three-body and the sequential calculations, respectively.}
\label{fig2}
\end{figure}

The fact that both calculations produce basically the same result is
encouraging. However, this good agreement is apparently inconsistent,
since it is achieved in spite of different energies of the lowest
2$^+_2$ resonance in $^{12}$C.  While in the full three-body calculation
this energy is 1.38 MeV, the sequential result is computed for an energy
of 1.75 MeV. In fact, when the sequential calculation is performed using
the same energy (1.38 MeV) and width (0.13 MeV) as in the full three-body 
calculation we obtain the rate given by the thick dotted curve in Fig.\ref{fig1}. 
As we can see, for the same parameters in the low-lying 2$^+_2$ resonance in
$^{12}$C, the sequential calculation gives a rate that, for temperatures beyond 2 GK,
is clearly above the one obtained in the three-body calculation.

For a better understanding it is useful to compare the $d{\cal B}/dE$
function for the $2^+\rightarrow 0^+_1$ transition obtained in the
full three-body calculation and in the sequential approximation from
Eq.(\ref{eq16}).  This comparison is shown in the upper part of
Fig.\ref{fig2} by the solid and dashed curves, respectively. The
$d{\cal B}/dE$ function in the sequential case is essentially a
lorentzian centered around the 1.75 MeV corresponding to the $2^+_2$
resonance energy. In the full three-body calculation the peak, as
expected, is located at about 1.4 MeV. Furthermore, a shoulder
appears for higher energies arising from the contribution from
continuum non-resonant states and the additional $2^+$ resonances
obtained in the three-body calculation. These contributions are not
included in the description in Ref.\cite{ang99}. 

However, due to the exponentials in Eq.(\ref{eq5}) or (\ref{eq11}), the high
energy region of the transition strength contributes very little to the reaction rate.
This is shown in the lower part of the figure, where we plot the integrands 
entering in the calculation of the reaction rate (Eqs.(\ref{eq5}) and (\ref{eq10})) as a function of the 
three-body kinetic energy for $T$=3, 5, 7, and 9 GK. The thick and thin curves 
correspond to the full three-body and the sequential calculations, respectively. As
expected, these integrands die rather fast with the energy, and the larger
the temperature the larger the energy range contributing to the reaction
rate.

As seen in the upper part of the figure, the full three-body
calculation accumulates part of the strength at smaller energies 
than the sequential calculation. 
This is first of all due to the different
resonance energies of 1.38~MeV in the three-body and 1.75~MeV in the
sequential calculation. 
For this reason, for low temperatures,
i.e. for $T\leq 3$~GK, where the integrand is vanishingly small for 
$E \gtrsim 2$ MeV (see the solid curves in the lower part of the figure), 
the three-body
calculation provides a higher reaction rate than in the sequential
approximation. The integral of the thick solid line in the lower part
of Fig.\ref{fig2} is almost twice the one of the thin solid
curve. When the temperature increases, the energy at which the
integrand begins to vanish also increases, and the sequential reaction
rate adds more strength compared to the three-body case.  As a
consequence both reaction rates approach each other, and they even
cross at some point. The integral of the thin dot-dashed curve in the
lower part of Fig.\ref{fig2} (sequential case for $T$=9 GK) is about
25\% bigger than the one of the thick dot-dashed curve (three-body
case for $T$=9 GK). Even for $T=9$ GK the integrand only allows
contribution up to energies around 5 MeV, which means that the
behavior of the $d{\cal B}/dE$ function for $E \gtrsim 5$ MeV is
rather unimportant.

Thus, from Fig.\ref{fig2} we can conclude that the full three-body calculation shifts part of the strength
to large energies, which contribute very little to the reaction rate. For 
this reason, for the same low lying 2$^+_2$ resonance energy in $^{12}$C, 
the full three-body calculation provides a smaller rate than the sequential
calculation. Or, in other words, for the three-body and sequential 
calculations to provide the same reaction rate, the 2$^+_2$ resonance
energy has to be smaller in the three-body case, in order to compensate 
the non-contributing part of the strength at large energies (for instance
1.38 MeV in the three-body calculation and 1.75 MeV in the sequential 
approach).

\paragraph*{The 2$^+_2$ resonance energy and the reaction rate.}

As seen in Fig.\ref{fig1}, for temperatures below about 3 GK the reaction rate of the triple $\alpha$
reaction is dominated by the $0^+\rightarrow 2^+_1$ transition, and the precise energy of the lowest
2$^+_2$ resonance in $^{12}$C does not play any role. Only for temperatures beyond 4 GK the energy of the
$2^+_2$ resonance has a sizable effect.

\begin{figure}[ht]
\psfig{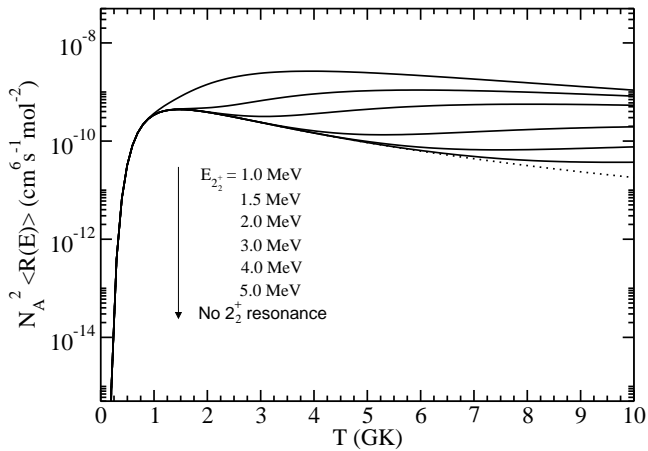}
\caption{Reaction rate in the sequential case for different energies of the lowest 2$^+_2$ resonance
in $^{12}$C. The energy increases from the upper curve to the lower one from 1 MeV up to 5 MeV. The dotted
curve is the calculation where the contribution from the $2^+\rightarrow 0^+_1$ transition has been 
completely removed.}
\label{fig3}
\end{figure}

Fig.\ref{fig3} shows the total sequential reaction rate for different
energies of the 2$^+_2$ resonance in $^{12}$C. The rate decreases
progressively when the resonance is placed at a higher and higher
energy. This happens because the higher the energy of the 2$^+_2$
resonance, the more strength located at high three-body energies,
where the integrand is small due to the exponential in
Eq.(\ref{eq5}) or (\ref{eq11}) (see the lower part of
Fig.\ref{fig2}). In fact, for a resonance energy of 5 MeV the
computed rate is quite similar to the one obtained when the resonance
is removed. Only for temperatures beyond around 7 GK some
difference appears.

\begin{figure}[ht]
\psfig{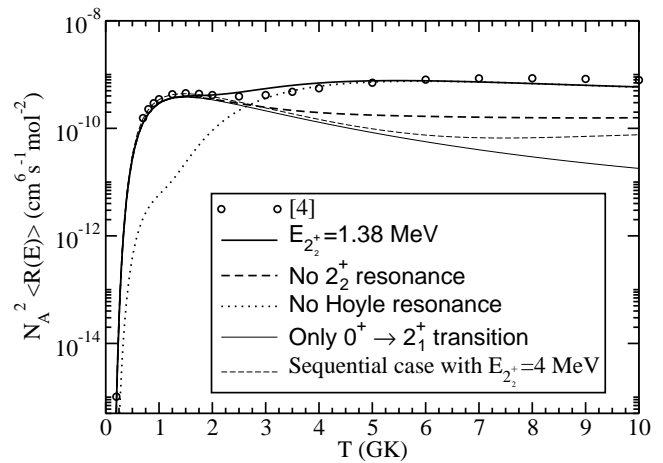}
\caption{Reaction rate for the three-body calculation when the $2^+_2$ resonance is placed at 1.38 MeV
(thick solid line), when the resonance is removed from the calculation (thick dashed line), when the full contribution
from the  $2^+\rightarrow 0^+_1$ transition is excluded (thin solid line), and when the Hoyle resonance
is removed (dotted line). The thin dashed curve is the calculation in the sequential case when the energy of the
$2^+_2$ resonance is 4.0 MeV. The open circles are taken from \cite{ang99}.}
\label{fig4}
\end{figure}

For the same reason, the same behavior is found for the three-body calculation. An increase
of the 2$^+_2$ resonance energy decreases the reaction rate. However,
full removal of the 2$^+_2$ resonance in $^{12}$C gives a rate that
differs quite a lot from the one obtained when the full contribution
from the $2^+\rightarrow 0^+_1$ transition is removed. This can be
seen in Fig.\ref{fig4}, where the thick solid line is the same calculation
as in Fig.\ref{fig1} (with the $2^+_2$ resonance at 1.38 MeV), the thick
dashed line is the calculation after removal of this 2$^+_2$ resonance,
and the thin solid line is the result when the full $2^+\rightarrow
0^+_1$ transition is excluded. The difference between the thin solid
line and the thick dashed line in the figure shows the importance of
including the additional continuum $2^+$ states in $^{12}$C. 

It is important
to keep in mind that the three-body calculation, together with the 2$^+_2$ resonance
at 1.38 MeV, includes as well all the other states obtained in \cite{alv07}, in particular
the 2$^+_3$ resonance around 4 MeV. For this reason, the thick dashed line in Fig.\ref{fig4},
where the resonance at 1.38 MeV has been suppressed, still contains the contribution
of the one at about 4 MeV, which is the one experimentally known. Then the thick dashed 
line in the figure gives a lower limit to the triple $\alpha$
reaction rate, since this is what we would get in case that the 2$^+$ resonance 
in $^{12}$C around 1.5-2.5 MeV is shown not to exist.

In this connection it is interesting to compare the thick dashed line
in the figure with the thin dashed line, which represents the reaction
rate obtained in the sequential approach when the lowest $2^+_2$
resonance in $^{12}$C is placed at 4.0 MeV. This is very close to the
energy of the lowest $2^+_2$ resonance fully confirmed
experimentally, and therefore this would be what provided by the sequential
calculation in case that the 2$^+$ resonance in $^{12}$C around 1.5-2.5 MeV does not
exist. The difference between both dashed curves shows the
effect of the non-resonant $2^+$ states in this case.

For illustration we also show in Fig.\ref{fig4} the reaction rate that
we obtain when the Hoyle resonance is removed from the calculation
(dotted curve in the figure). As we can see, this state is responsible
for an increase of the reaction of about two orders of magnitude for
temperatures below 3 GK.

Thus, from Fig.\ref{fig4} we can conclude that in case that the low-lying
2$^+_2$ resonance is confirmed not to exist (such that the lowest one is the
already known resonance close to 4 MeV above threshold) the sequential
approach would underestimate the triple-$\alpha$ reaction rate. The non-resonant
2$^+$ states in $^{12}$C are enough to increase significantly the rate at
high temperatures.

\paragraph*{Summary and conclusions.}

We investigate the reaction rate for the triple $\alpha$ reaction. We
used the full three-body calculation described in \cite{die10} and the
sequential description of \cite{ang99}.  For the same set of
three-body resonances we find that the full three-body calculation
gives smaller rates than the sequential approximation for temperatures
above $\sim 3$ GK.

We focus on the importance of the experimentally uncertain lowest
$2^+_2$ resonance in $^{12}$C which to a large extent determines the
reaction rate for temperatures above 3~GK.  Different theoretical
calculations predict a $2^+_2$ resonance in $^{12}$C at around 1.0-2.5
MeV. We find that when the full and sequential calculations are
performed taking a $2^+_2$ resonance energy of 1.38~MeV and 1.75~MeV, respectively,
the computed total reaction rates are quite similar for the whole range 
temperature. The only exception is for temperatures ranging between 
about 2 and 5 GK, where the full calculation gives a rate slightly above the
one obtained with the sequential calculation.  The reason for 
this behavior is that part of the strength in the
three-body case is moved to higher energies due to the higher 2$^+$
resonances and non-resonant continuum 2$^+$ states.  The high energy
part of this strength contributes very little, and the missing
low-energy strength has to be compensated by a lower-lying $2^+$
resonance. For the same energy and width of the 2$^+_2$ resonance, the full 
three-body calculation contributes less than the sequential approximation.

If the lowest $2^+_2$ resonance in $^{12}$C by chance should be at about
4 MeV, where the lowest established $2^+$ state is located, then the
reaction rate is higher for the full three-body calculation than
obtained from the sequential approximation.  This is due to
contributions from the non-resonant continuum states in the three-body
calculation and the insignificant contribution from a high-lying
resonance in the sequential calculation.  The numerical result can be
taken as a lower limit to the reaction rate, which for high
temperatures can be up to about one order of magnitude above the
result obtained in the sequential picture.

Summarizing, a detailed description of the triple $\alpha$ reaction
rate for temperatures beyond 3 GK requires a careful treatment of the
$2^+\rightarrow 0^+_1$ transition. The non-resonant continuum $2^+$
states, not included in the sequential description, are very
important.  For a given energy of the lowest $2^+_2$ resonance in $^{12}$C below 
about $3$~MeV, a sequential description of the process
overestimates the reaction rate. For a larger resonance energy the rate 
of three-body calculation exceeds that of the sequential approximation due to the
contribution of the non-resonant continuum states.

\paragraph*{Acknowledgments.}
This work was partly supported by funds provided by DGI of MEC (Spain)
under contract No.  FIS2008-01301. One of us (R.D.) acknowledges
support by a Ph.D. I3P grant from CSIC and the European Social Fund.

\end{document}